# Ferroelectric switching at symmetry-broken interfaces by local control of dislocation networks

Laurent Molino[1], Leena Aggarwal[1], Vladimir Enaldiev[3], Ryan Plumadore[1], Vladimir Falko[2,3,4*], Adina Luican-Mayer[1**]

[1] Department of Physics, University of Ottawa, Ottawa, Canada
[2] National Graphene Institute, University of Manchester, Manchester, UK
[3] Department of Physics and Astronomy, University of Manchester, Manchester, UK
[4] Henry Royce Institute for Advanced Materials, University of Manchester, Manchester, UK
* vladimir.falko@manchester.ac.uk
** luican-mayer@uottawa.ca

**Abstract**
Semiconducting ferroelectric materials with low energy polarisation switching offer a platform for next-generation electronics such as ferroelectric field-effect transistors. Ferroelectric domains at symmetry-broken interfaces of transition metal dichalcogenide films provide an opportunity to combine the potential of semiconducting ferroelectrics with the design flexibility of two-dimensional material devices. Here, local control of ferroelectric domains in a marginally twisted WS$_2$ bilayer is demonstrated with a scanning tunneling microscope at room temperature, and their observed reversible evolution understood using a string-like model of the domain wall network. We identify two characteristic regimes of domain evolution: (i) elastic bending of partial screw dislocations separating smaller domains with twin stacking and (ii) formation of perfect screw dislocations by merging pairs of primary domain walls. We also show that the latter act as the seeds for the reversible restoration of the inverted polarisation. These results open the possibility to achieve full control over atomically thin semiconducting ferroelectric domains using local electric fields, which is a critical step towards their technological use.

**Main**
Overcoming challenges in modern computer engineering and telecommunication technologies, requires compact multi-functional components. Atomically thin films of semiconducting transition metal dichalcogenides (TMDs) offer multiple advantages for such a development. They retain robust semiconducting properties down to sub-nanometer thickness, which allows for an efficient transistor operation, and they exhibit strong light-matter coupling. Moreover, TMDs have recently emerged as ferroelectric two-dimensional (2D) materials, in which switching between two twin stacking orders produces room-temperature stable out-of-plane polarisation with low switching barriers[1–5]. Semiconducting ferroelectrics with low energy barrier polarisation switching are of interest, as they enable ferroelectric field-effect transistors, devices that combine memory storage and logic processing into a single device[6]. The possibility of assembling vertical stacks of designer sequences from atomically thin planes of layered van der Waal materials allows for unprecedented flexibility in engineering novel electronic devices[7,8]. In particular, previously inexistent electronic properties can emerge in these systems through relative rotation of the atomically thin layers, which leads to the formation of moiré patterns and reconstruction domains[9–11].

Lack of inversion symmetry among individual monolayers of TMDs permits control over the symmetry of interfaces formed in layer-by-layer assembled heterostructures. For example, when assembling two layers with inverted orientation of unit cells, one obtains bilayers with a local inversion-symmetric structure, which reconstructs[10,12] (at small twist angles) into a honeycomb set of 2H stacking domains[12–14], separated by domain walls which have the form of intra-layer shear solitons[15]. Moreover, the assembly of layers with parallel orientation of unit cells (P-bilayers) generates structures with non-symmetric interfaces which allow for a spontaneous interlayer charge transfer and, therefore, out-of-plane ferroelectric polarisation[1–4,16]. The bilayers with parallel orientation of monolayer unit cells exhibit two energetically equivalent favourable stacking orders, which are mirror images of each other (illustrated in Fig. 1a). In one type of stacking (XM'), the metal site M' of the bottom layer appears under the chalcogen site X of the top layer, whereas the chalcogen/metal sites (X'/M) in the bottom/top layers sit under/over the middle of the honeycomb of the other layer's lattice. In the second type of stacking (MX'), the metal site M of the top layer appears over the chalcogen site X' of the bottom layer, whereas the chalcogen/metal sites (X/M') in the top/bottom layers sit over/under the middle of the honeycomb of the other layer's lattice. For small-angle twisted P-bilayers, in-plane relaxation[10,12–14] leads to the alternating twinning of XM' and MX' domains carrying opposite electric polarisation[12–14,16,17]. These domains are separated by domain walls which are similar to partial dislocations in the 3R polytype of bulk TMDs.

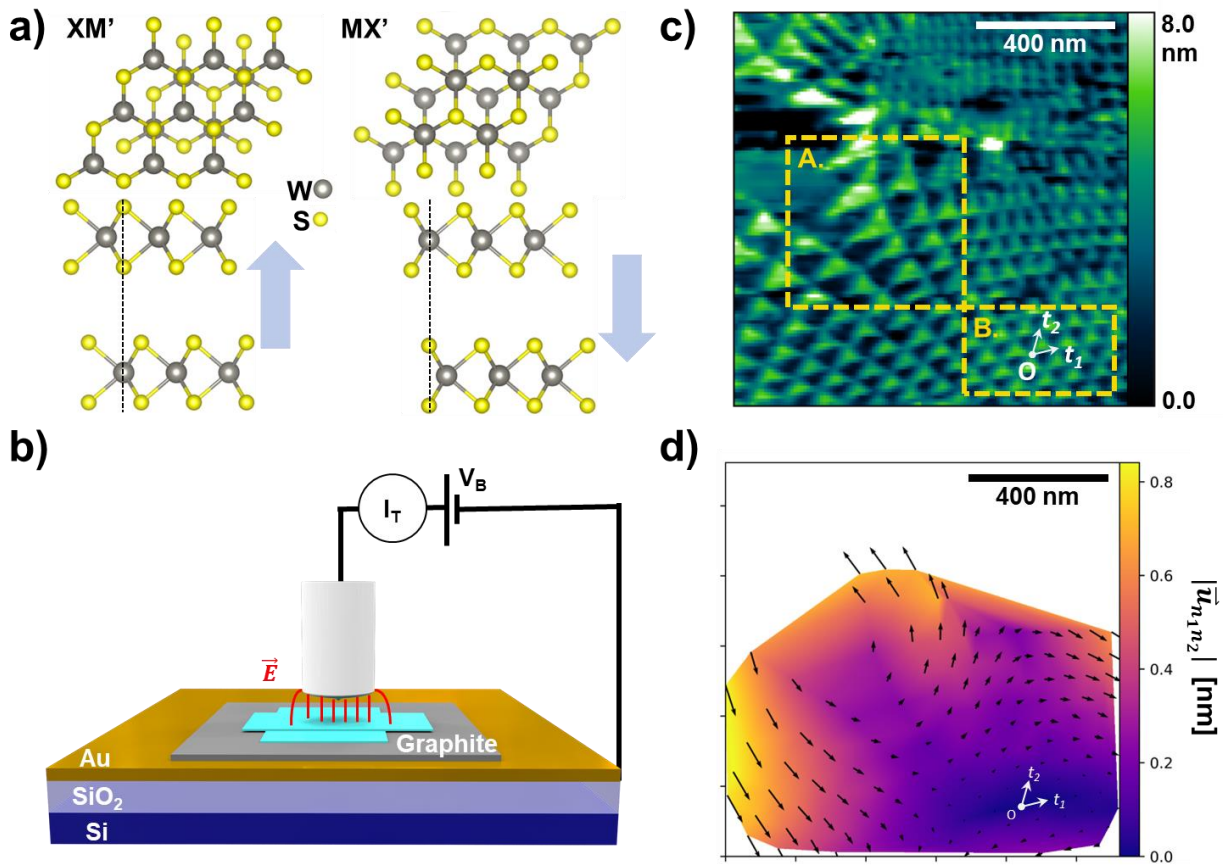

*Figure 1: a) Top and side views of XM' and MX' stackings, with arrows indicating the direction of electric polarisation. b) Schematic of the STM experiment, indicating the electric field produced by applying a bias voltage between the tip and the sample. c) Large STM topographic map ($V_B$ = -1.3 V, $I_T$ = 50pA), showing two distinct regions in the sample indicated by A and B. In area B, the triangular moiré superstructure is homogeneous, with equilateral triangles, and has periodicity given by the translation vectors $\vec{t}_{1,2}$. In area A, the moiré pattern is skewed due to a small strain in one of the layers (whose effect is magnified by the moiré superlattice effect[20]). d) Spatial displacement field map for the inter-layer deformations, obtained using Eq. (1) from the variation of moiré superlattice pattern in c).*

Here, two P-aligned $WS_2$ samples were assembled as schematically indicated in Fig. 1b (optical micrographs and fabrication details are available in Supplementary Figure 1 in SI). A thin graphite film was placed under the TMDs to improve electrical contact. Scanning tunneling microscopy and spectroscopy (STM/STS) was used to image and control the reconstruction domains (XM' and MX') formed in the twisted $WS_2$ samples. As indicated in Fig. 1b, such an experiment implies the presence of a perpendicular electric field between the tip and the sample, which can couple to the local polarisation of the XM' and MX' stacking areas. A typical scanning tunneling topograph for our samples (Fig. 1c) shows large triangular domains of XM' and MX' stacking (Fig. 1a), separated by a network of domain walls with clearly identifiable nodes (to which one can attribute[12] energetically unfavourable XX' stacking characterised by vertical alignment of chalcogens in the top and bottom layers). The formation of these domains is understood by the fact that in twisted P-aligned bilayers of hexagonal TMDs the relaxation of a moiré pattern into domains occurs for the twist angles below 2.5°[13,14,16].

Within the triangular domains of the network in Fig. 1c, we first focus on the area B containing a network of approximately equilateral triangular domains. These equilateral domains have periodicity of approximately 78 nm corresponding[13,14,16,18] to a twist angle 0.23°. This area will be used as a reference for the analysis of a larger-scale aperiodic domain pattern (domains of various sizes and with pronounced anisotropy). Such a distortion of the domain network is caused by a small accidental strain in either of the assembled $WS_2$ monolayers, inflicted in the fabrication process[19]. Theoretically, it has been demonstrated[20] that a moiré pattern acts as a magnifying glass for small deformations in the constituent monolayers of the bilayer. Strain-induced shifts of the atomic registry between the layer – even as small as a fraction of the lattice constant – lead to a shift of the equivalent stacking areas in the moiré pattern by distances comparable to the moiré superlattice period. Here, we use this generic property of moiré patterns to map the displacement field distribution (due to the accidental strain) and analyse the position of domain wall network nodes, which feature XX' stacking. For this, we successively enumerate all nodes by integers $(n_1, n_2)$, starting from the referenced point O= (0,0) inside area B in Fig. 1c with approximately equilateral domains. For this area, we determine a twist angle $\theta = \frac{|\vec{a}_1|}{|\vec{t}_1|} \approx 0.0041$ (≈0.23°) from the moiré period, $\vec{t}_1$ in area B ($|\vec{t}_{1,2}| \approx 78$ nm nm and $|\vec{a}_{1,2}| = 0.315$ nm). Then, we take into account that, without strain, the positions, $\vec{r}_{n_1,n_2}$, of the XX' network nodes in the rest of the sample would correspond to the mutual shift of atomic registry such that $n_1\vec{a}_1 + n_2\vec{a}_2 = \theta\hat{z} \times \vec{r}_{n_1,n_2}$, whereas the difference between the actual, $\vec{R}_{n_1,n_2}$, and expected, $\vec{r}_{n_1,n_2}$, positions of the network sites enables us to determine the strain-induced interlayer shift (superimposed onto the local lattice reconstruction into domains) of the monolayer $WS_2$ lattices as[20],

$$\vec{u}(\vec{R}_{n_1,n_2}) = \theta\hat{z} \times \vec{R}_{n_1,n_2} - n_1\vec{a}_1 - n_2\vec{a}_2. \quad (1)$$

The resulting deformations map is shown in Fig. 1d, indicating a shear strain vortex in top left corner and discernible deformations from the sides of the analyzed area.

The effect of the reconstruction domains on the local electronic properties, as measured using STM/STS, is discussed in Fig. 2. The STM topographic map in Fig. 2a shows a typical region with triangular domains. By observing the domain wall bending (domain expansion/shrinking) in response to the applied biased voltage, the bright and dark domains were identified from the ferroelectric polarisation as MX' and XM' stacking regions, respectively[21]. The contrast in the STM topography of the two stackings is consistent with previously reported STM measurements of twisted TMD homobilayers with similar scanning parameters[22]: likely, due to the prominence of the metal atom in this stacking. The tunneling spectrum in Fig. 2b, averaged over all domains, shows the conduction and valence band edges at approximately 0.6 eV and -1.5 eV respectively, indicating an electronic band gap consistent with previous measurements of this system[23]. The differential tunneling conductance (dI/dV) maps acquired at the indicated voltages in Fig. 2c, demonstrate the dependence of the local density of states on the stacking order inside the domains and at the domain walls.

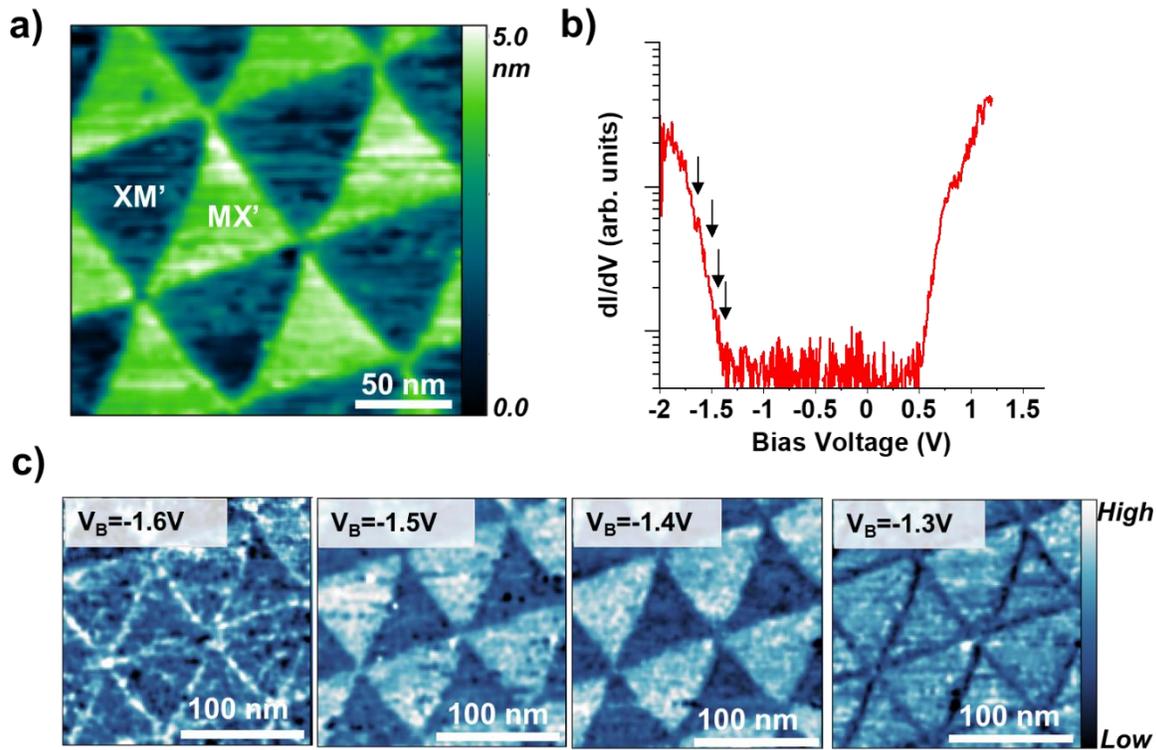

*Figure 2: a) STM topographic map of a typical region with equilateral triangular stacking order domains ($I_t$= 100 pA, $V_B$ = -1.3V). b) STS averaged over bright and dark domains (set point: $I_T$ = 60 pA and $V_B$ = -1.9 V). c) dI/dV maps at the indicated bias voltages ($I_t$= 100pA).*

Although a scanning tunneling microscope uses an atomically sharp tip end as a local probe, the overall shape of the tip and its holder make the electric field between the tip and the sample approximately uniform over a typical domain size (Fig. 1b). To verify this, we compared forward and reverse topographic scans (Supplementary Figure 2 in SI) which are found to be identical, supporting the scenario of an approximately uniform out-of-plane electric field. Such a perpendicular electric field can be controlled by varying the bias voltage between the tip and the sample (Fig. 1b). In this particular case, an out-of-plane electric field lifts the energetic equivalence of the XM' and MX' domains, expanding areas of those which have favoured direction of polarisation, thus, curving domain walls.[10] In Fig. 3a we experimentally demonstrate that for a sufficiently large bias voltage, reversing the direction of the perpendicular electric field resulted in significant shape changes of the domain walls, while the nodes of the domains remained in the same position. This is due to the topological nature of domain walls, which precludes nucleation of new domains, with the node density determined by the local twist angle. All changes in the domain pattern are produced by the displacements of existing domain walls[16], whose ends remain pinned to the nodes of the initial domain wall network. In the data presented in Fig. 3a we find that the brighter domains tend to shrink for the negative bias voltage at the expense of the darker domains; in contrast, for a positive bias voltage, the darker domains shrink at the expense of the brighter domains, who tend to enlarge. We quantitatively analyze the shapes of such equilateral triangular domains using a previously developed model[1] for the calibration of the electric displacement field between tip and sample to the applied bias voltage. We will subsequently use the resulting calibration to capture the behaviour of domain redistribution by vertical electric field across a complex network of elongated domains (e.g., in area A of Fig. 1c).

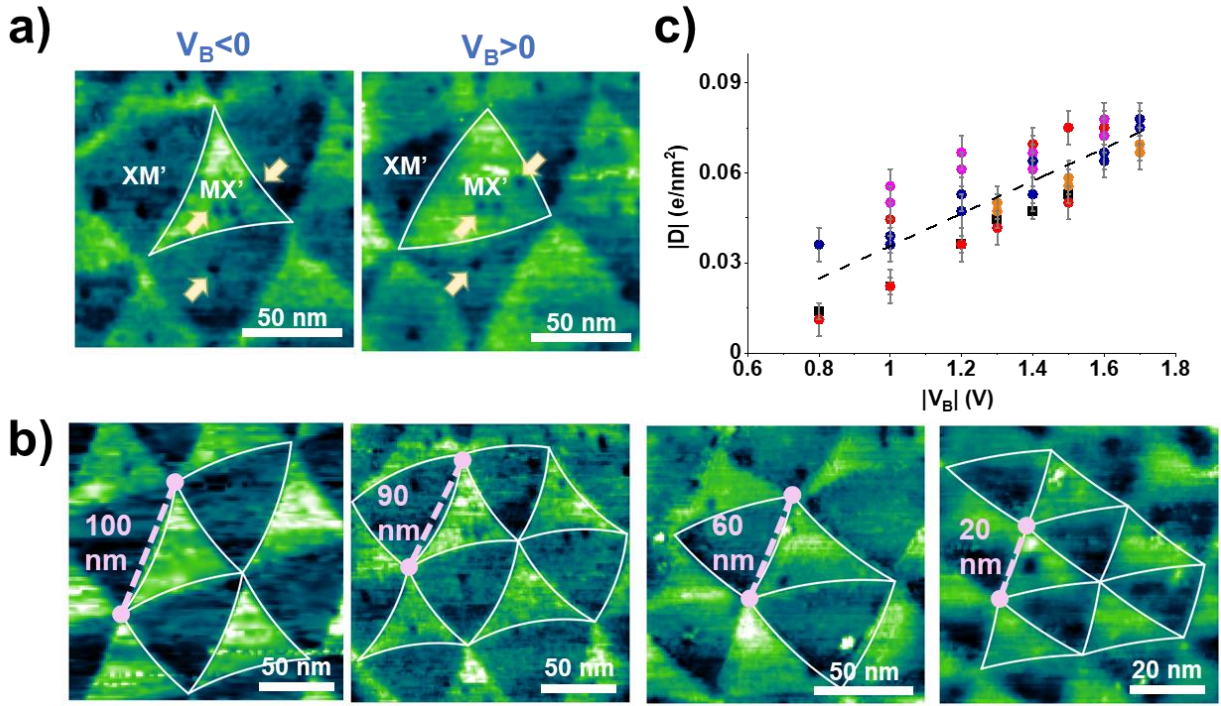

*Figure 3: a)* STM topographic maps acquired at $I_T= 100$ pA and $V_B =\pm 1.3$ V. The white lines represent fits with string-like model using as parameter the electric displacement field values $D=-0.075$ e/nm$^2$, $0.050$ e/nm$^2$. We note the presence of defects (indicated by arrows), which served as a marker for the areas of interest[24,25]. The domain wall dynamics was realised in an imperfect crystal, where the domain walls were not pinned by defects, and where defects did not act as nucleation sites. In addition, the defects were found stable, even when crossed by a moving domain wall. *b)* STM topographic maps of different areas with domains of indicated lengths ($I_T= 100$ pA

and $V_B = -1.4$ V). The white lines represent fits with the equilateral triangle domain wall model (from left to right: D=-0.047 e/nm², -0.070 e/nm², -0.053 e/nm², -0.08 e/nm²). **c)** The electric displacement field as a function of bias voltage for five areas. Colours indicate different areas, squares correspond to Sample 1 Tip 1, circles correspond to Sample 2 Tip 2 (SI).

For the equilateral domains, experimentally accessible electric fields lead only to a small bending of the domain walls (partial dislocations) with ends clamped at the domain wall network nodes. Using a previously developed model[1,4], the shape of each domain wall can be described by its transverse deflection, $y(x) = \int_0^x \frac{f(x')}{\sqrt{1-f^2(x')}} dx'$, from a straight line, $0 \leq x \leq \ell$, connecting a pair of consecutive nodes. In this expression, the function $f(x')$ is a root of a cubic polynomial, $f^3 - Af - B\left(x - \frac{\ell}{2}\right) = 0$ where $A = \frac{w}{\tilde{w}} + 2 \approx 3.52$, with $w = 1.05$ eV/nm characterising minimal single domain wall energy per its unit length (realised for armchair direction in WS$_2$) and $\tilde{w} = 0.69$ eV/nm taking into account orientation-dependent parts (energy of partial dislocations is maximal for zigzag direction). The value of electric displacement field, $D$, is incorporated into another parameter in the solution, $B = \frac{2D\Delta}{\tilde{w}}$ where $\Delta = 62$ mV is the double-layer potential drop caused by the ferroelectric polarisation in XM' (MX') domains[3].

Using this model, we analyzed STM topographic maps (e.g., such as in area B in Fig. 1b) that showed almost equilateral triangular domains at different voltages and at different length scales. By adjusting the value of the electric displacement field $D$ for each analyzed domain wall, we obtained the curves represented by white lines in Fig. 3a and Fig. 3b. We note that this model adequately captures the shape of domain bending at different electric fields (positive and negative) as well as at different lengths between the nodes of the network. Fig. 3c summarizes the values obtained for the electric displacement field as a function of bias voltage by performing such fits of domain wall shape for five areas on two different samples and with two different tips. The relationship (see Fig. 3c) between the bias voltage $V_B$ and the electric displacement field, $D = \frac{\varepsilon_0 V_B}{d^*}$, gives us an effective distance, $d^*=1.0 \pm 0.7$ nm, between the tip and the sample, which is close to a typical tip-sample separation in the reported STM experiment.

As we have pointed out earlier, even a small strain in one of the layers can distort the equilateral domains into elongated ones, such as those indicated in region A area of Fig. 1c. Examples of such measured domains are highlighted in Fig. 4b, where – for a strong enough electric field $D>D^*$ – the long stretches of the curved domain walls (which structurally are the same a partial screw dislocations) merge together into a 'full' perfect screw dislocation (PSD), highlighted by red lines in Fig. 4b. These PSDs separate pairs of identically polarised MX' domains. Hence, we extend the domain network modelling to deformed triangular domains, where two of the inter-node distances, $\ell_{1,2}$, are longer than the third one, $\ell_3$. For the non-equilateral domains, the anisotropy of DW energy is described by $w + \tilde{w} \sin^2\left(\varphi_{1,2,3} + \arctan(y'_{1,2,3})\right)$, where angles $\varphi_{1,2,3}$ characterise deviations of DWs from the closest armchair direction at $D=0$, (see SI), and $\arctan(y'_i)$ appears due to a local transversal deflection, $y_i$ induced by electric field. The DW shapes are found by minimizing the energy functional elaborated in detail in SI with the following boundary conditions: $y_{1,2}\left(\mathcal{L}\cos(\alpha_{1,2})\right) = \mathcal{L}\sin(\alpha_{1,2})$ and $y_{1,2,3}(\ell_{1,2,3}) = y_3(0) = 0$. The first of those accounts for the emergence of a PSD streak of length $\mathcal{L}$ (see SI), where we also define angles $\alpha_{1,2}$. To find these PSD parameters, we take into account that the transversal shifts $y_1$ and $y_2$ bring the DW together at the PSD end, $x_{1,2} = \mathcal{L} \cos(\alpha_{1,2})$. For $\alpha_{1,2} \ll 1$, justified for the strongly elongated domains in Fig. 4a, the condition can be approximated by,

$$\mathcal{L} = \frac{\ell_1 - \sqrt{\ell_1 \frac{2A_1}{B} \tan(\alpha_1)}}{\cos(\alpha_1)} = \frac{\ell_2 - \sqrt{\ell_2 \frac{2A_2}{B} \tan(\alpha_2)}}{\cos(\alpha_2)}, \quad \alpha_1 + \alpha_2 = \arccos\left(\frac{\ell_1^2 + \ell_2^2 - \ell_3^2}{2\ell_1 \ell_2}\right), \tag{2}$$

where $A_{1,2} = A - 3\sin^2\varphi_{1,2}$. Equations (2) enable us to describe all the details of the field-induced changes of the domain wall network, as marked on the maps in Fig. 4. For a quantitative comparison, in Fig. 4c we compare the theoretically computed lengths $\mathcal{L}$ of the PSD streaks with the measured values $L_{\text{exp}}$ for different domains and electric fields.

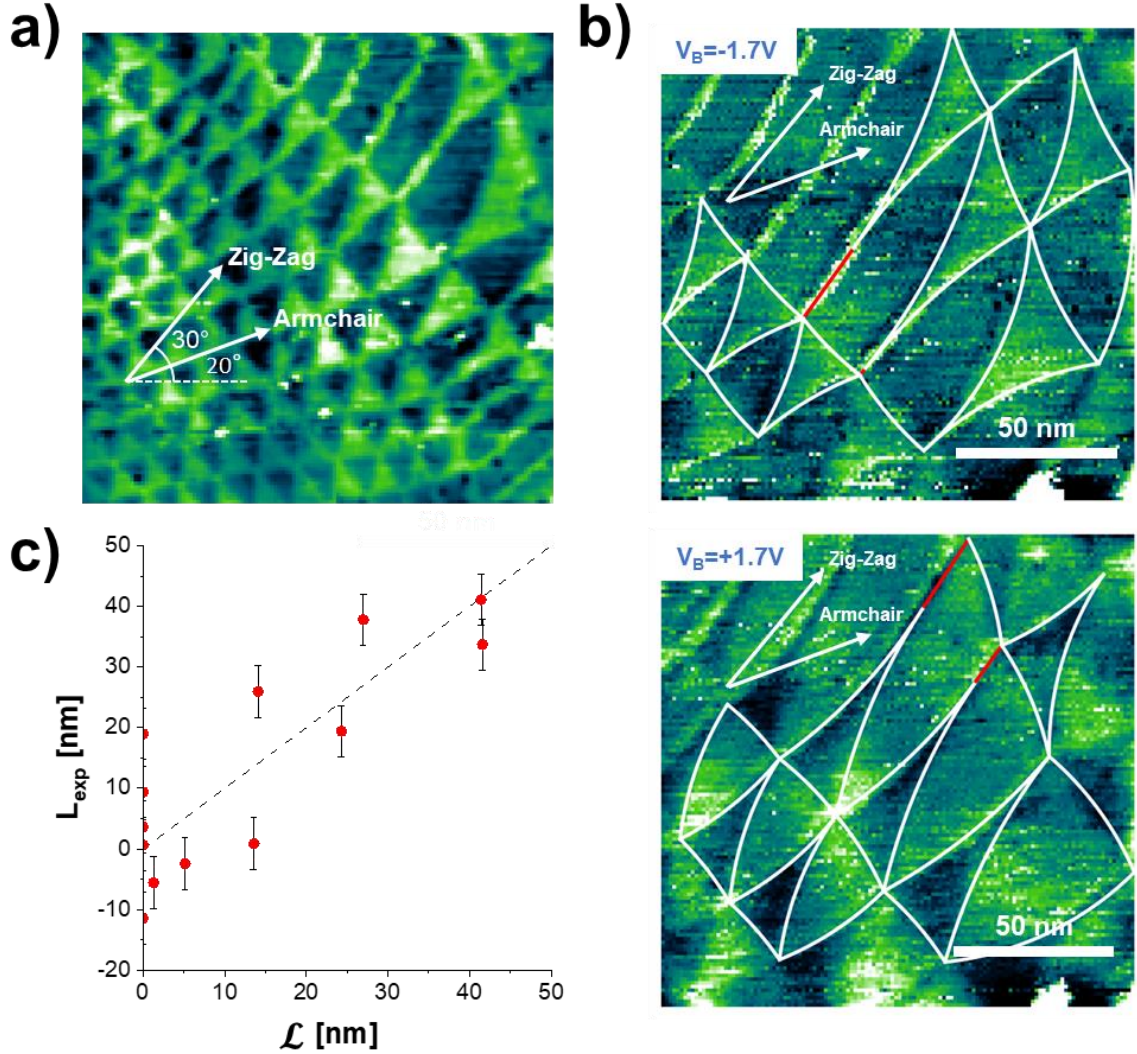

*Figure 4: a) STM topography with observed angle of armchair direction indicated. Data acquired at $V_B$ = -1.3 V, $I_T$ = 50pA. b) STM topography of a region in top-left of panel a). Images taken at $I_T$= 100 pA and $V_B$ = ± 1.7 V. Single domain wall fits from string-like models plotted in white, predicted double domain walls plotted in red with $A = A(\varphi)$ where $\varphi$ is the angle of the triangle side relative to the nearest armchair direction. D values of -0.067 e/nm² and 0.070 e/nm² are used for the two images. As we already determined the effective distance that characterized the equilateral triangles (Fig. 3), in the case of the elongated triangles, drawing the shapes of the domain walls was done free of tunable parameters: the only input is the position of the network nodes. c) Predicted PSD length $\mathcal{L}$ compared to observed length for elongated domains (see SI for details on estimating $L_{exp}$) The error bars reflect the pixel size of our images.*

To summarize, we have demonstrated that for each marginally twisted samples the analysis of the ferroelectric domain network can give detailed information about the deformations in the assembled layers. These deformations lead to the variation of domain sizes across the network, producing a variety of switching conditions for individual domains. We demonstrate that one can achieve full local control over the domain network, by squeezing out the unfavorable polarization from the larger and elongated domains. This happens through merging pairs of partial dislocations into a perfect dislocation. Note that, despite an apparent disappearance of the unfavorably polarised domains, the surface polarisation remains reversible, as, upon the reversed bias, perfect 'full' dislocations split up into pairs of partial dislocation, nucleating domains of opposite stacking order. We have thus demonstrated reversible local control of ferroelectric dipole moments in a transition metal dichalcogenide homobilayer.

**Methods:**
The sample was assembled using the tear-and-stack technique[26,27]. $WS_2$ and graphite crystals were purchased from HQ Graphene and Graphenea respectively and were mechanically exfoliated onto $Si/SiO_2$ substrates. Monolayer $WS_2$ was identified optically, and the thickness was confirmed with atomic force microscopy (AFM). Suitable graphite flakes were identified and were picked up with a polydimethylsiloxane (PDMS) stamp covered with Polypropylene carbonate (PPC) film on a glass slide. The graphite on the polymer stamp was then used to pick up monolayer $WS_2$. The two layers of $WS_2$ correspond to two halves of a large flake, allowing for precise rotational control. The PPC film was then cut and the 2D material stack was transferred onto a $Si/SiO_2$ substrate coated with Ti/Au (5nm/150nm). The sample was then annealed under ultra-high vacuum at 300°C for several hours. The surface was cleaned using an AFM tip[28,29]. STM measurements were performed at pressures below $10^{-9}$ Torr, at room temperature. For STS we used bias modulation of 5.0mV, 1.325 kHz.

**Acknowledgements:**
V.F. and V.E. acknowledge funding from: EC-FET European Graphene Flagship Core3 Project, EPSRC grants EP/S030719/1 and EP/V007033/1, Lloyd Register Foundation Nanotechnology Grant. AL-M, L.M., L.A., R.P. acknowledge funding from: Natural Sciences and Engineering Research Council of Canada (NSERC) Discovery Grant RGPIN-2022-05215, Ontario Early Researcher Award ER-16-218 and NSERC Strategic Project QC2DM STPGP 521420.

# Ferroelectric switching at symmetry-broken interfaces by local control of dislocation networks

Laurent Molino[1], Leena Aggarwal[1], Vladimir Enaldiev[3], Ryan Plumadore[1], Vladimir Falko[2,3,4*], Adina Luican-Mayer[1**]

[1] Department of Physics, University of Ottawa, Ottawa, Canada
[2] National Graphene Institute, University of Manchester, Manchester, UK
[3] Department of Physics and Astronomy, University of Manchester, Manchester, UK
[4] Henry Royce Institute for Advanced Materials, University of Manchester, Manchester, UK
* vladimir.falko@manchester.ac.uk
** luican-mayer@uottawa.ca

## A. Experimental Details

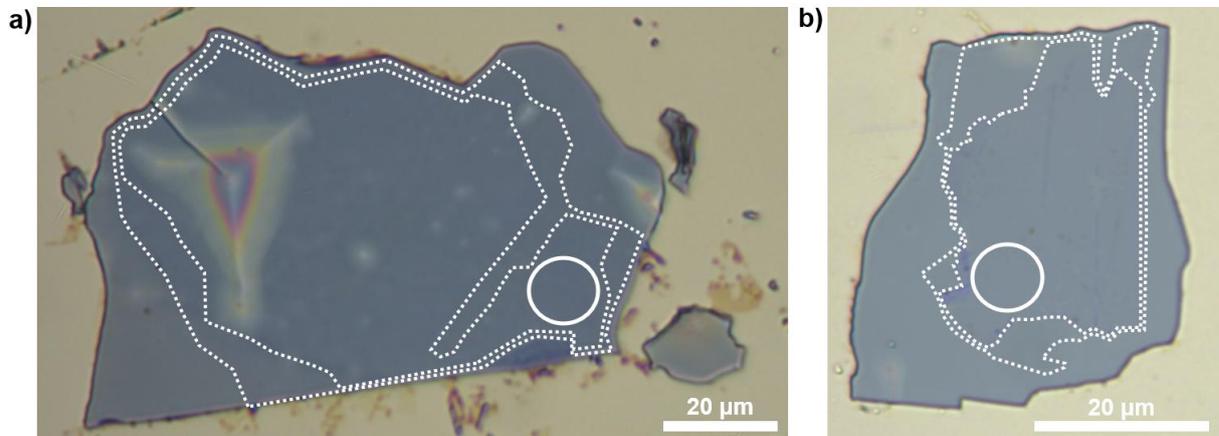

**Supplementary Figure 1.** Optical micrograph of the Samples 1 (a) and 2 (b). Dotted lines indicate approximate boundaries of the $WS_2$ layers. Circles indicate approximate area where data were collected.

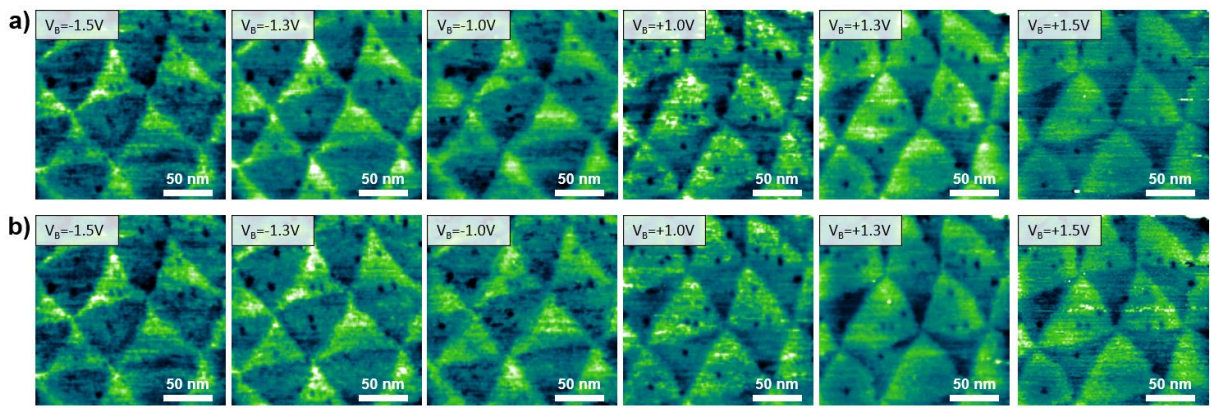

**Supplementary Figure 2.** STM topographic maps in forward (a) and reverse (b) scanning directions. Data acquired at $I_T = 100$ pA and indicated biases.

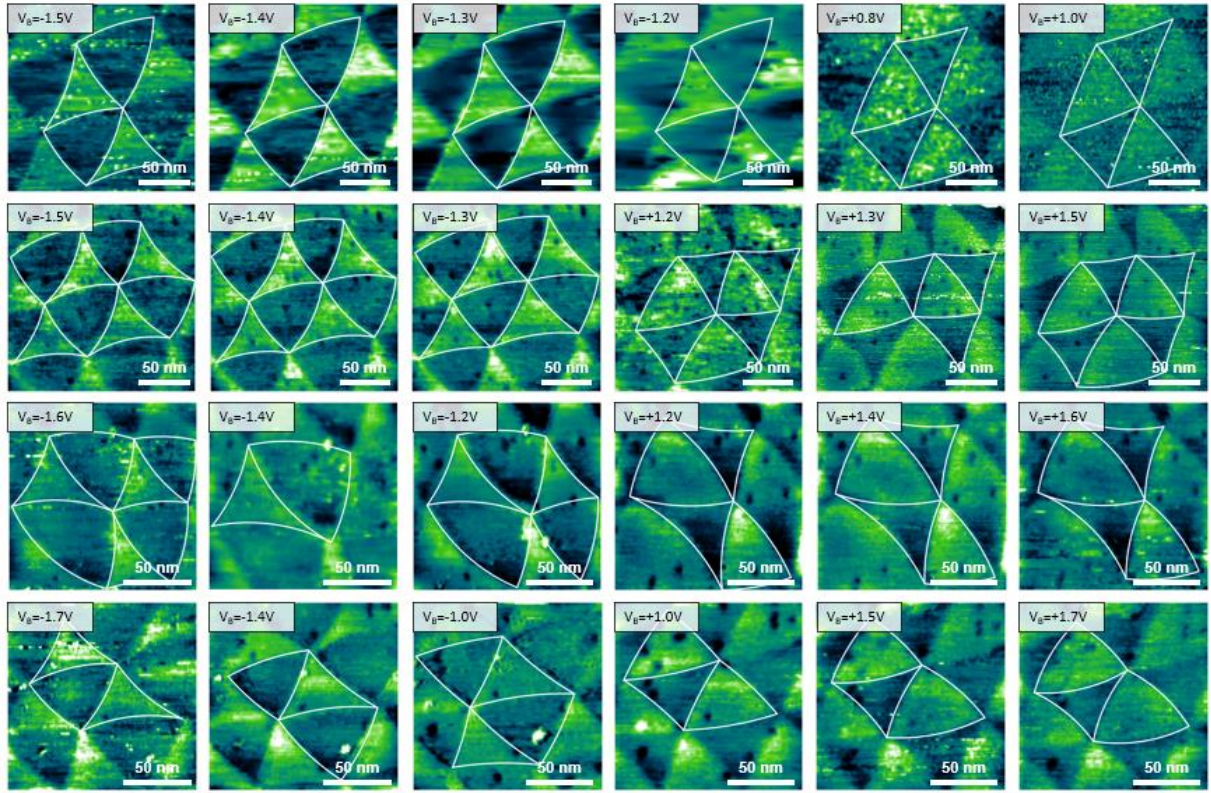

**Supplementary Figure 3.** STM topography and domain wall fits. Data collected on Sample 1 with tip 1 (a), and Sample 2 with tip 2 (b-d). Data acquired at $I_T$ = 100 pA and indicated biases.

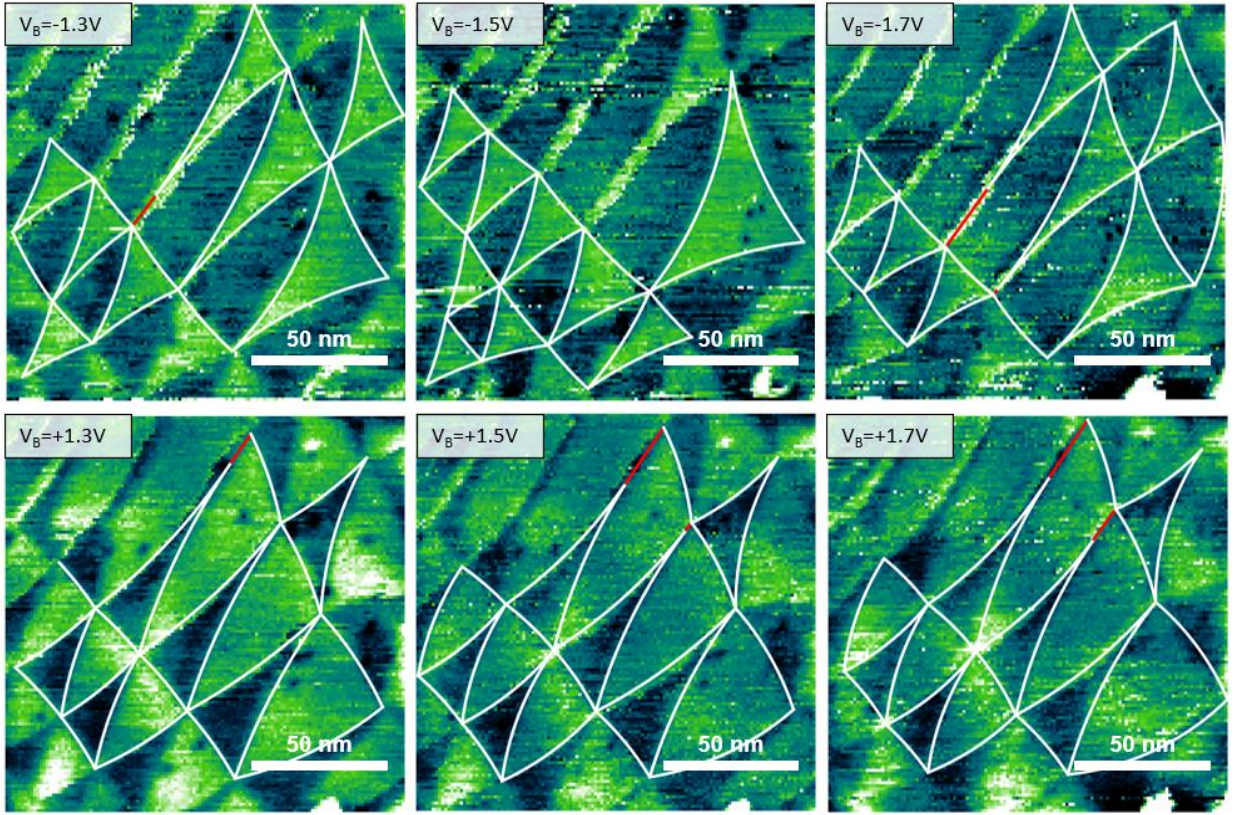

**Supplementary Figure 4.** Elongated triangle domain wall fits. Calculated double domain wall shown in red. Data acquired on Sample 2 with tip 2 at $I_T = 100$ pA and indicated biases.

**B. Extension of network model for elongated domains.**

To describe merging of two domain walls (DWs) for elongated domains we modify the model used for equilateral domains[1]. We define an elongated domain as the one having a short DW with internode distance $\ell_3$, and two much longer DWs characterised by internode distance $\ell_{1,2} \gg \ell_3$. To characterise bending energy of each DW we use an expression established earlier[1], $w + \widetilde{w} \sin^2(\Phi_{1,2,3})$, where $w = 1.05 \, eV/nm$ is DW energy per unit length in armchair direction, $\widetilde{w} = 0.69 \, eV/nm$ is an orientation-dependent part of the energy, and $\Phi_{1,2,3}$ are angles defining local orientation of the DW with respect to the armchair direction. In the model we also take into account opportunity for the two long DWs to merge into a perfect screw dislocation oriented along zigzag direction with energy per unit length $u = 2.45 \, eV/nm$. To describe bending of long DWs we introduce functions, $y_{1,2}(x_{1,2})$ determined in the interval $\mathcal{L} \cos(\alpha_{1,2}) \leq x_{1,2} \leq \ell_{1,2}$, while shape of the short DW is determined by $y_3(x_3)$ in the interval $0 \leq x_3 \leq \ell_3$. Then, local orientation of the DWs is $\Phi_i = \varphi_i + \arctan(y'_i)$, $i = 1,2,3$, and $\varphi_i$ are angles between direction of the DW in zero electric field and the closest armchair crystallographic axis (see Figure S5). To find $y_{1,2,3}(x_{1,2,3})$, $\mathcal{L}$, and $\alpha_{1,2}$ we minimise the following functional

$$\mathcal{E}[y_1(x_1), y_2(x_2)] = \sum_{i=1,2,3} \int_{\delta_i}^{\ell_i} \left[ \left( w + \widetilde{w} \frac{(\sin(\varphi_i) + y'_i \cos(\varphi_i))^2}{1 + y'^2_i} \right) \sqrt{1 + y'^2_i} - 2D\Delta y_i \right] dx_i + \quad (S1)$$

$$+ \left[ \mathcal{L}u - \frac{\mathcal{L}^2 D\Delta}{4} \sum_{i=1,2} \sin(2\alpha_i) \right].$$

Here, the first term in the integrand describes the orientation-dependent energy of partial dislocations due to their bending ($\sin(\Phi_i) = \frac{\sin(\varphi_i) + y'_i \cos(\varphi_i)}{\sqrt{1+y'^2_i}}$), the second term stands for the energy gain ($-2D\Delta$) from increasing area of a domain with the stacking promoted by the out-of-plane electric field, $D/\epsilon_0$, which couples to the ferroelectric polarisation density, $\epsilon_0 \Delta$, determined by interlayer voltage drop $\Delta = 62$ mV for 3R WS$_2$ bilayers ($\epsilon_0$ is vacuum permittivity); integration limits are $\delta_{1,2} = \mathcal{L}\cos(\alpha_{1,2})$, $\delta_3 = 0$. Terms outside of the integral allow for merging of the long DWs with formation of the perfect screw dislocation of $\mathcal{L}$-length projected into the intervals $0 < x_{1,2} < \mathcal{L}\cos(\alpha_{1,2})$. For long DWs this leads to boundary conditions (BCs): $y_{1,2}(\mathcal{L}\cos(\alpha_{1,2})) = \mathcal{L}\sin(\alpha_{1,2})$, $y_{1,2}(\ell_{1,2}) = 0$, in which $\mathcal{L}$ and $\alpha_{1,2}$ are additional variation parameters that to be found from minimisation of (S1). However, for short DW BCs are: $y_3(\ell_3) = y_3(0) = 0$.

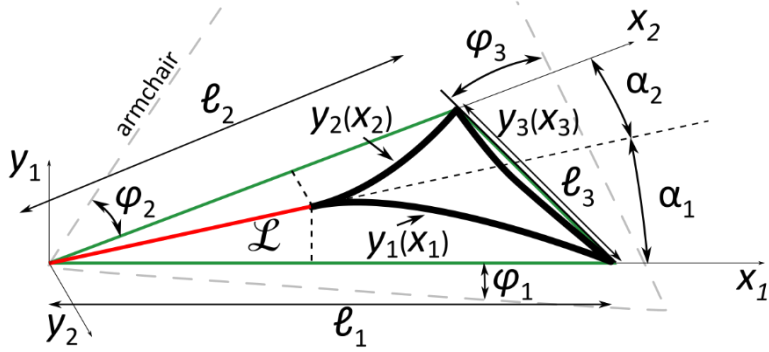

**Supplementary Figure 5**. Green triangle shows DWs (partial dislocations) surrounding a single elongated 3R domain at zero electric field. Bold black lines show shape of partial dislocations under out-of-plane electric field. Red straight segment is streak of perfect screw dislocation formed by merging of the two long DWs. Dashed gray lines show armchair directions for which DW energy is minimal.

General solutions of the corresponding Euler-Lagrange equations are expressed via roots of the algebraic equations:

$$\frac{y'^3_i}{(1+y'^2_i)^{3/2}} - \frac{\frac{w}{\tilde{w}}+2-3\sin^2\varphi_i}{\cos 2\varphi_i} \frac{y'_i}{\sqrt{1+y'^2_i}} - \frac{\tan\varphi_i}{(1+y'^2_i)^{\frac{3}{2}}} - \frac{2D\Delta}{\tilde{w}\cos 2\varphi_i}x + C_1 = 0, \quad (S2)$$

where $C_1$ is integration constant. In the limit $\varphi_{1,2,3} = 0$, which corresponds to equilateral domains, solution of (S2) is reduced to that derived in Reference 1, noting that $\frac{f_i}{\sqrt{1-f_i^2}} = y'_i$ and choosing $C_1 = D\Delta\ell$, which ensures symmetric shape of the bended DWs. Here we solve (S2) in the limit $y'^2_i \ll 1$ which is relevant for experimentally observed elongated domains. In this case in (S2) we leave only terms $\propto y'_i$, which results in parabolic shapes for DWs:

$$y_{1,2}(x_{1,2}) = \frac{D\Delta/\tilde{w}}{\frac{w}{\tilde{w}}+2-3\sin^2\varphi_{1,2}} (\ell_{1,2} - x_{1,2})\left(x_{1,2} - \frac{\mathcal{L}^2 \cos^2(\alpha_{1,2})}{\ell_{1,2}}\right), \quad (S3)$$

$$y_3(x_3) = \frac{D\Delta/\tilde{w}}{\frac{w}{\tilde{w}}+2-3\sin^2\varphi_3} (\ell_3 - x_3)x_3. \quad (S4)$$

In Equation (S3), parameters of PSD are determined by touching condition of the DWs at $x_{1,2} = \mathcal{L} \cos(\alpha_{1,2})$, which leads to minimum of functional (S1):

$$\begin{cases} \tan(\alpha_1) = \dfrac{D\Delta/\widetilde{w}}{\frac{W}{\widetilde{w}}+2-3\sin^2\varphi_1} \dfrac{(\mathcal{L}\cos(\alpha_1)-\ell_1)^2}{\ell_1}, \\ \tan(\alpha_2) = \dfrac{D\Delta/\widetilde{w}}{\frac{W}{\widetilde{w}}+2-3\sin^2\varphi_2} \dfrac{(\mathcal{L}\cos(\alpha_2)-\ell_2)^2}{\ell_2}, \\ \alpha_1 + \alpha_2 = \arccos\left(\dfrac{\ell_1^2+\ell_2^2-\ell_3^2}{2\ell_1\ell_2}\right). \end{cases} \quad (S5)$$

To solve (S4) we exclude $\mathcal{L}$ with the help of the first two equations and, then, using the third equation determine $\alpha_{1,2}$. With known $\alpha_{1,2}$ length of the PSD streak reads as:

$$\mathcal{L} = \dfrac{\ell_1-\sqrt{\ell_1\dfrac{\left(\frac{W}{\widetilde{w}}+2-3\sin^2\varphi_1\right)}{D\Delta/\widetilde{w}}\tan(\alpha_1)}}{\cos(\alpha_1)} = \dfrac{\ell_2-\sqrt{\ell_2\dfrac{\left(\frac{W}{\widetilde{w}}+2-3\sin^2\varphi_2\right)}{D\Delta/\widetilde{w}}\tan(\alpha_2)}}{\cos(\alpha_2)}. \quad (S6)$$

**How we obtain the experimental PDS length $L_{exp}$.**

In Fig. 4c in the main text, experimental PSD length $L_{exp}$ is plotted against predicted PSD length $\mathcal{L}$. The theoretically predicted separation between the partial screw dislocations that run close to each other just before they reach the merging point is comparable to our image pixel resolution which was ~1.6nm. Therefore, in our STM images, what we observe as merged length is an overestimation of the PSD length by an amount $L_{res}$ (Supplementary Figure 6), which can be calculated in each case from Equation (S6) taking into account the pixel size. To account for this, when calculating $L_{exp}$ reported in Fig. 4c, for each domain, the measured length of the merged domain wall in the STM image, $L_{STM}$ (Supplementary Figure 6) was reduced by $L_{res}$, $L_{exp}=L_{STM}-L_{res}$.

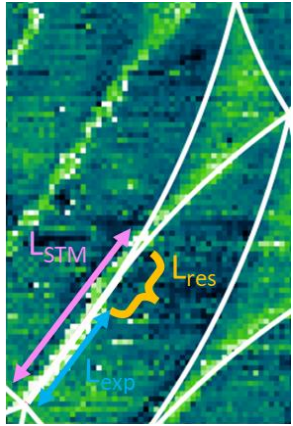

**Supplementary Figure 6.** How we obtain the experimental PDS length $L_{exp}$.

We note that in some elongated triangles, at some biases, no merging is predicted to occur: a domain node is reached before the partial screw dislocations can merge.

## C. Computation of displacements and strain components

To obtain map of displacements presented in Figure 1d of the main text, we identified nodes of the domain wall network and label them successively by pair of integers, see Figure S6a, with origin in the point O. Vectors $\vec{t}_1 = \vec{R}_{1,0}$ and $\vec{t}_2$ (60° rotated $\vec{t}_1$) form basis of moiré superlattice in area with equilateral domains determining armchair axes in the layers and twist angle $\theta = |\vec{a}_1|/|\vec{t}_1|$. As it was explained in the main text, displacement at node position $\vec{R}_{n_1,n_2}$ is expressed as follows

$$\vec{u}(\vec{R}_{n_1,n_2}) = \theta \hat{z} \times \vec{R}_{n_1,n_2} - n_1 \vec{a}_1 - n_2 \vec{a}_2, \qquad (S7)$$

where primary lattice vectors of WS$_2$ monolayer are orthogonal to the superlattice vectors, $\vec{a}_{1,2} = \theta[\hat{z} \times \vec{t}_{1,2}]$.

Having determined displacement field, we compute strain tensor components for each triangle spanned by three nodes $(n_1, n_2)$, $(n_1 + 1, n_2)$, $(n_1, n_2 + 1)$ using first order Taylor expansion of the displacements $\vec{u}(\vec{R}_{n_1+1,n_2})$ and $\vec{u}(\vec{R}_{n_1,n_2+1})$:

$$\begin{cases} \vec{u}(\vec{R}_{n_1+1,n_2}) - \vec{u}(\vec{R}_{n_1,n_2}) = (\vec{R}_{n_1+1,n_2} - \vec{R}_{n_1,n_2}) \cdot \left( \frac{\partial \vec{u}}{\partial x}\Big|_{\vec{R}_{n_1,n_2}}, \frac{\partial \vec{u}}{\partial y}\Big|_{\vec{R}_{n_1,n_2}} \right) \\ \vec{u}(\vec{R}_{n_1,n_2+1}) - \vec{u}(\vec{R}_{n_1,n_2}) = (\vec{R}_{n_1,n_2+1} - \vec{R}_{n_1,n_2}) \cdot \left( \frac{\partial \vec{u}}{\partial x}\Big|_{\vec{R}_{n_1,n_2}}, \frac{\partial \vec{u}}{\partial y}\Big|_{\vec{R}_{n_1,n_2}} \right). \end{cases} \qquad (S8)$$

Then, we diagonalise local strain tensor $u_{ij}\big|_{\vec{R}_{n_1,n_2}} = \frac{1}{2}\left(\frac{\partial u_i}{\partial x_j} + \frac{\partial u_j}{\partial x_i}\right)\Big|_{\vec{R}_{n_1,n_2}}$ making appropriate rotation $C(\alpha)$, by angle $\alpha = \frac{1}{2}\arctan\left(\frac{2u_{xy}}{u_{xx}-u_{yy}}\right)$. In the rotated frame strain tensor $\tilde{u}_{ij}\big|_{\vec{R}_{n_1,n_2}} = C^{-1}(\alpha) u_{ij}\big|_{\vec{R}_{n_1,n_2}} C(\alpha) = diag(u_h + u_s, u_h - u_s)$ is diagonal and characterised by hydrostatic, $u_h = \frac{u_{xx}+u_{yy}}{2}$, and shear, $u_s = \sqrt{\left(\frac{u_{xx}-u_{yy}}{2}\right)^2 + u_{xy}^2}$, components. In Figure S6(b-e) we demonstrate distributions for all extracted quantities $\vec{u}(\vec{R}_{n_1,n_2})$, $u_h(\vec{R}_{n_1,n_2})$, $u_s(\vec{R}_{n_1,n_2})$ and $\alpha(\vec{R}_{n_1,n_2})$ characterising the local deformations.

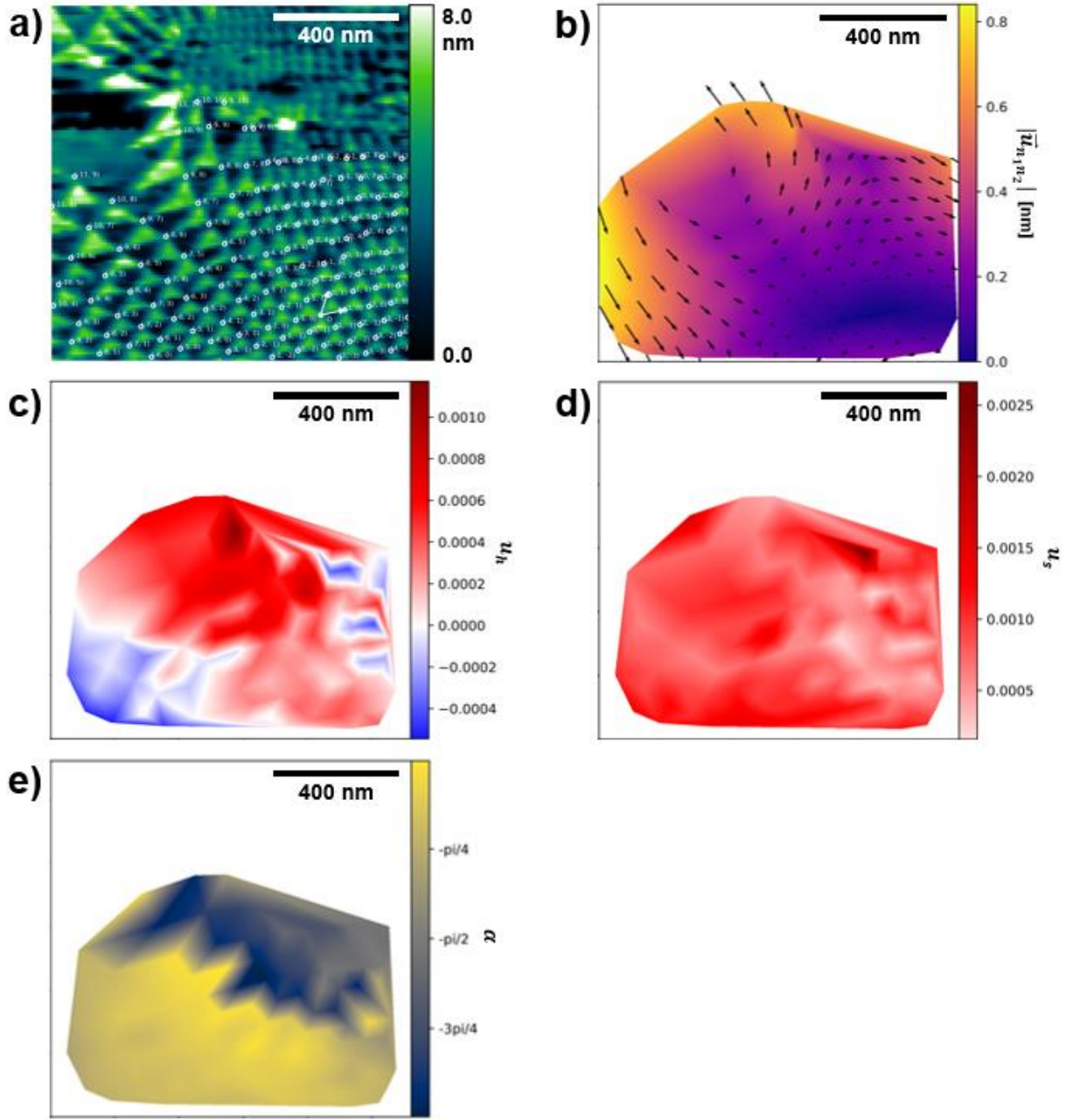

**Supplementary Figure 7**. a) STM topographic map ($V_B$ = -1.3 V, $I_T$ = 50pA), with vertices label with their indices. Point O and vectors $\vec{t}_{1,2}$ are indicated. b) Displacement magnitude and direction corresponding to panel c). c) Hydrostatic strain. d) Shear strain. e) Rotation angle $\alpha$.